\documentclass[a4paper,fleqn]{cas-dc}

\usepackage[sort&compress,numbers]{natbib}
\usepackage[utf8]{inputenc}
\usepackage{textcomp}
\DeclareUnicodeCharacter{22EF}{\dots}
\DeclareUnicodeCharacter{0302}{\^}
\DeclareUnicodeCharacter{0301}{\'}
\usepackage{soul}
\usepackage{ulem}

\usepackage{siunitx}
\usepackage[version=4]{mhchem}

\def\tsc#1{\csdef{#1}{\textsc{\lowercase{#1}}\xspace}}
\tsc{WGM}
\tsc{QE}
\tsc{EP}
\tsc{PMS}
\tsc{BEC}
\tsc{DE}

\begin{document}
\let\WriteBookmarks\relax
\def\floatpagepagefraction{1}
\def\textpagefraction{.001}
\shorttitle{Athos-Graphene a new 2D Carbon Allotrope}
\shortauthors{Lima et~al.}

\title [mode = title]{Athos-Graphene: Computational Discovery of an Art-Inspired 2D Carbon Anode for Lithium-Ion Batteries}

\author[1]{Kleuton A. L. Lima}
\affiliation[1]{
organization={Department of Applied Physics and Center for Computational Engineering and Sciences},
addressline={State University of Campinas}, 
city={Campinas},
postcode={13083-859}, 
state={SP},
country={Brazil}}
\credit{Conceptualization of this study, Methodology, Review and editing, Investigation, Formal analysis, Writing -- review \& editing, Writing -- original draft}
                 
\author[2]{José A. S. Laranjeira}
\affiliation[2]{
organization={Modeling and Molecular Simulation Group},
addressline={São Paulo State University (UNESP), School of Sciences}, 
city={Bauru},
postcode={17033-360}, 
state={SP},
country={Brazil}}
\credit{Conceptualization of this study, Methodology, Review and editing, Investigation, Formal analysis, Writing -- review \& editing, Writing -- original draft}

\author[2]{Nicolas F. Martins}
\credit{Conceptualization of this study, Methodology, Review and editing, Investigation, Formal analysis, Writing -- review \& editing, Writing -- original draft}

\author[2]{Julio R. Sambrano}
\credit{Conceptualization of this study, Methodology, Review and editing, Investigation, Formal analysis, Writing -- review \& editing, Writing -- original draft}

\author[3,4]{Alexandre C. Dias}
\affiliation[3]{
organization={Institute of Physics},
addressline={University of Brasília}, 
city={Brasília },
postcode={70910‑900}, 
state={DF},
country={Brazil}}
\affiliation[4]{
organization={International Center of Physics},
addressline={University of Brasília}, 
city={Brasília },
postcode={70910‑900}, 
state={DF},
country={Brazil}}
\credit{Conceptualization of this study, Methodology, Review and editing, Investigation, Formal analysis, Writing -- review \& editing, Writing -- original draft}

\author[1]{Douglas S. Galvão}
\credit{Conceptualization of this study, Methodology, Review and editing, Investigation, Formal analysis, Writing -- review \& editing, Writing -- original draft}

\author[5]{Luiz A. Ribeiro Junior}
\cormark[1]
\cortext[cor1]{Corresponding author}
\affiliation[3]{
organization={Computational Materials Laboratory, LCCMat, Institute of Physics},
addressline={University of Brasília}, 
city={Brasília },
postcode={70910‑900}, 
state={DF},
country={Brazil}}
\credit{Supervision, Funding Acquisition, Review and editing, Formal analysis, Writing -- review \& editing, Writing -- original draft}

\begin{abstract}
Two-dimensional (2D) carbon allotropes have attracted growing interest for their structural versatility and potential in energy storage and nanoelectronics. We propose Athos-Graphene (AG), a novel 2D carbon allotrope inspired by the geometric patterns of Brazilian artist Athos Bulcão. Designed using density functional theory, AG features a periodic structure with high thermodynamic and thermal stability, as evidenced by a low cohesive energy (\SI{-7.96}{\electronvolt/atom}), the absence of imaginary phonon modes, and robust performance in \textit{ab initio} molecular dynamics simulations up to \SI{1000}{\kelvin}. It exhibits anisotropic mechanical properties, with Young’s modulus values of \SI{585}{\giga\pascal} and \SI{600}{\giga\pascal}along the $x$- and $y$-directions, and Poisson's ratios of \SI{0.19}{} and \SI{0.17}{}, respectively. Electronic structure analyses confirm its metallic behavior, while optical studies reveal anisotropic absorption in the visible and UV regions. For lithium-ion storage, Athos-Graphene shows strong \ce{Li} adsorption (\SIrange{-2.3}{-1.0}{\electronvolt}), a high theoretical capacity of \SI{836.78}{mAh/g}, and a low average open-circuit voltage of \SI{0.54}{\volt}. Lithium diffusion barriers are as low as \SI{0.3}{\electronvolt} on the surface and \SI{0.66}{\electronvolt} between layers, with a high diffusion coefficient ($>6 \times 10^{–6}$ \si{\square\cm/s}). These features highlight AG as a promising anode material for high-performance lithium-ion batteries.
\end{abstract}




\maketitle

\section{Introduction}

Two-dimensional (2D) carbon-based allotropes have become one of the major topics in materials science due to their exceptional physical and chemical properties, particularly their high electrical conductivity, mechanical strength, and surface tunability \cite{tiwari2016magical,hirsch2010era,enyashin2011graphene}. Among them, graphene has been of fundamental importance for a new era in carbon-based research. Its unique electronic and mechanical properties have been widely exploited in many applications, including energy storage ones \cite{li2018graphene,zhu2014graphene,brownson2011overview}. However, graphene's dense hexagonal lattice limits accessible surface area and ion adsorption capacity, with a ratio of 6 carbon atoms to one ion, making it less effective for high-capacity lithium-ion storage \cite{chen2015structural}. Porous 2D carbon allotropes with non-hexagonal rings can offer improved ion access, greater specific surface areas, and superior electrochemical performance. These structural features aid lithium storage, making such materials promising for advanced lithium-ion battery (LIB) anodes. \cite{zhao2015graphite,wang2016porous,han2014porous,xu2012porous,xing2019porous,zhang2019recent}.

Despite the rapidly growing number of computationally proposed 2D carbon allotropes \cite{jana2021emerging,enyashin2011graphene,lusk2009creation,zhang2019art}, not all exhibit metallic behavior, a crucial requirement to minimize internal resistance and enhance charge transport in LIB anodes \cite{rajkamal2019carbon}. Many predicted structures are semiconducting or have high diffusion barriers and insufficient lithium-ion adsorption capabilities, limiting their practical use as anode materials \cite{jana2021emerging,rajkamal2019carbon,qi2017nanostructured,mahmood2016nanostructured}.

In this context, porous 2D carbon lattices, especially those incorporating non-hexagonal rings, have attracted attention due to their ability to combine structural stability with enhanced ion accessibility, increased specific surface area, fast diffusion, and high storage capacity \cite{hwang2013multilayer,jang2013graphdiyne,xiao2011hierarchically}. The rational design of such materials requires structural novelty and potential for bottom-up synthesis guided by well-defined molecular motifs.

\begin{figure*}[!htb]
    \centering
    \includegraphics[width=\linewidth]{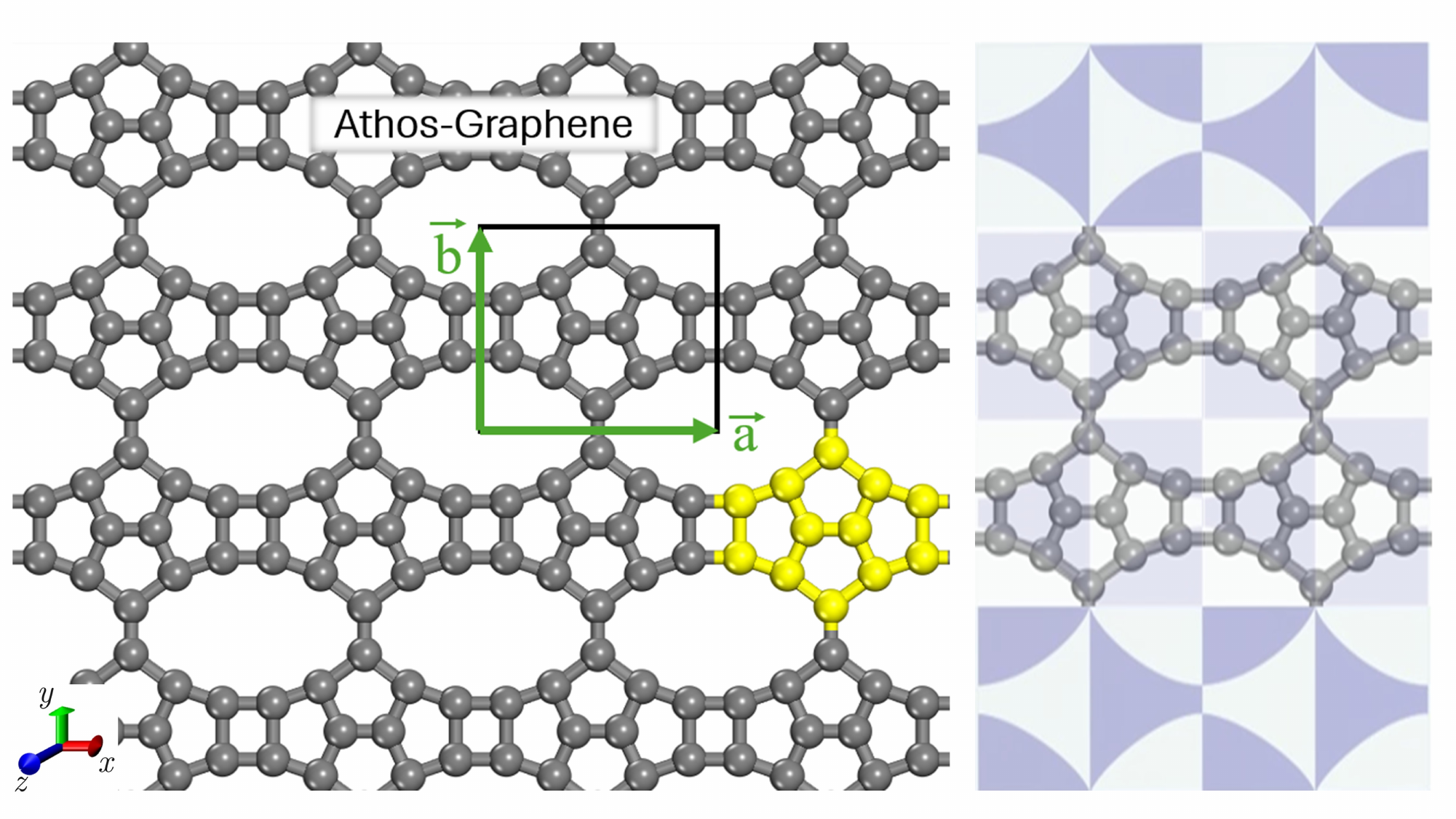}
    \caption{Schematic representation of the atomic structure of Athos-Graphene (AG). The highlighted yellow atoms represent the chemical precursor Dicyclobuta[de,ij]naphthalene. The unit cell is highlighted in black, with its related lattice vectors $\vec{a}$ and $\vec{b}$. The right panel illustrates one of the Bulcão's tiling mosaics, which inspired the AG structure.}
    \label{fig:system}
\end{figure*}

Here, to address some of these limitations, we propose a novel computationally designed 2D carbon allotrope, named Athos-Graphene (AG), see Fig.~\ref{fig:system}. The AG geometry is based on the molecular motif Dicyclobuta[de,ij]naphthalene, a polycyclic conjugated hydrocarbon with fused cyclobutadiene and naphthalene rings \cite{macaluso2004dicyclobuta}. The structural topology of this molecular structure offers a promising modular assembly strategy through established synthetic chemistry pathways, making ATG a good candidate for future experimental realization.

The name AG is a tribute to the Brazilian artist Athos Bulcão \cite{wanderley2006azulejo}, renowned for incorporating geometric modularity and symmetry into architectural tile patterns. The interplay between modular order and controlled variation in Bulcão's works is a conceptual parallel to the atomic arrangement in this 2D lattice, illustrating the harmony between artistic design and materials engineering. Similar inspiration gave rise to penta-graphene, a 2D carbon allotrope that resembles the Cairo pentagonal tiling \cite{zhang2015penta}.

This work presents the structural design and density functional theory (DFT) AG characterization. We explore its structural and thermal stability, mechanical behavior, electronic and optical properties, and performance as a lithium-ion storage material through adsorption, diffusion, and open-circuit voltage analyses. The results show that AG is a multifunctional 2D material with potential nanoelectronics and energy storage systems applications.

\section{Methodology}

The AG structural, electronic, and lithium-ion storage properties were investigated using DFT simulations using the CASTEP computational package \cite{clark2005first}. All calculations were carried out with the Perdew-Burke-Ernzerhof (PBE) exchange-correlation functional within the generalized gradient approximation (GGA) framework \cite{perdew1996generalized}. A plane-wave energy cutoff of \SI{450}{\electronvolt} and a Monkhorst-Pack \textbf{k}-point mesh of $10\times 10\times 1$ were adopted for Brillouin zone integration. Structural optimization was performed under periodic boundary conditions, with convergence criteria set to residual forces below \SI{0.001}{\electronvolt/\angstrom} and residual stress below \SI{0.01}{\giga\pascal}. A vacuum spacing of \SI{20}{\angstrom} was introduced along the out-of-plane direction to eliminate spurious interlayer interactions. Long-range dispersion forces were accounted for using the Grimme D2 semi-empirical van der Waals correction scheme \cite{grimme2006semiempirical}.

Phonon dispersion calculations were performed using DFT perturbation theory (DFPT) \cite{baroni2001phonons} to assess the structural dynamical stability. In addition, the thermal stability was evaluated through \textit{ab initio} molecular dynamics (AIMD) simulations performed at \SI{1000}{\kelvin} for a duration of \SI{5}{\ps}, using a Nosé-Hoover thermostat \cite{nose1984unified} in a canonical ensemble (NVT). These simulations used a $2\times 2 \times 1$ supercell and a $5\times 5\times 1$ \textbf{k}-points mesh. Structural integrity was monitored throughout the simulation to identify potential bond breakages, structural reconstructions, or structural failures from thermal effects.

Electronic band structures and projected density of states (PDOS) were computed using a denser \textbf{k}-point grids of $10\times 10\times 1$ and $20\times 20\times 1$, respectively. The metallic behavior was confirmed through analysis of band dispersion and electron localization function (ELF) maps.

The lithium adsorption properties were explored by placing \ce{Li} atoms at various high-symmetry and randomly selected adsorption sites on the AG surface, followed by full structural relaxation to determine the most energetically favorable configurations, using the adsorption locator tool \cite{metropolis1953equation,kirkpatrick1983optimization,vcerny1985thermodynamical}.

The adsorption energy (E$_{\text{ads}}$) was calculated using the expression: $E_{\text{ads}} = E_{\text{AG+Li}} - \left( E_{\text{AG}} + E_{\text{Li}} \right)$, where $E_{\text{AG+Li}}$ is the AG total energy with a \ce{Li} atom adsorbed, $E_{\text{AG}}$ is the energy of the AG pristine monolayer, and $E_{\text{Li}}$ corresponds to the energy of an isolated \ce{Li} atom in vacuum.

The ion diffusion barriers were determined using the CI-NEB method (Climbing Image Nudged Elastic Band) \cite{makri2019preconditioning,barzilai1988two,bitzek2006structural}. This analysis provided energy profiles for \ce{Li} migration pathways and information on AG and interlayer mobility, which are critical for assessing the performance of the material as a lithium-ion battery anode.

\section{Results}

\subsection{Structural and mechanical properties}

The AG atomic structure is presented in Fig.~\ref{fig:system}, which presents a planar 2D carbon lattice composed of sp$^2$ hybridized carbon atoms arranged in a geometrically rich tiling. In contrast to the uniform hexagonal motif of pristine graphene, AG integrates a diverse set of polygonal rings, namely four-, five-, and twelve-membered units, organized in a topologically non-trivial periodic framework that results in a porous configuration. To evaluate the structural stability of AG, the cohesive energy (\( E_{\text{coh}} \)), was calculated as \( E_{\text{coh}} = (E_{\text{AG}} - n E_{\text{C}})/n \), where \( E_{\text{AG}} \) is the total energy of the AG monolayer, \( E_{\text{C}} \) is the energy per carbon atom and \( n \) is the number of carbon atoms in the unit cell. The obtained \( E_{\text{coh}} \) value was of \SI{-7.96}{\electronvolt}, which is same range of other 2D carbon allotropes values \cite{jana2021emerging,enyashin2011graphene,lusk2009creation,zhang2019art}. 

\begin{figure}[!htb]
    \centering
    \includegraphics[width=\linewidth]{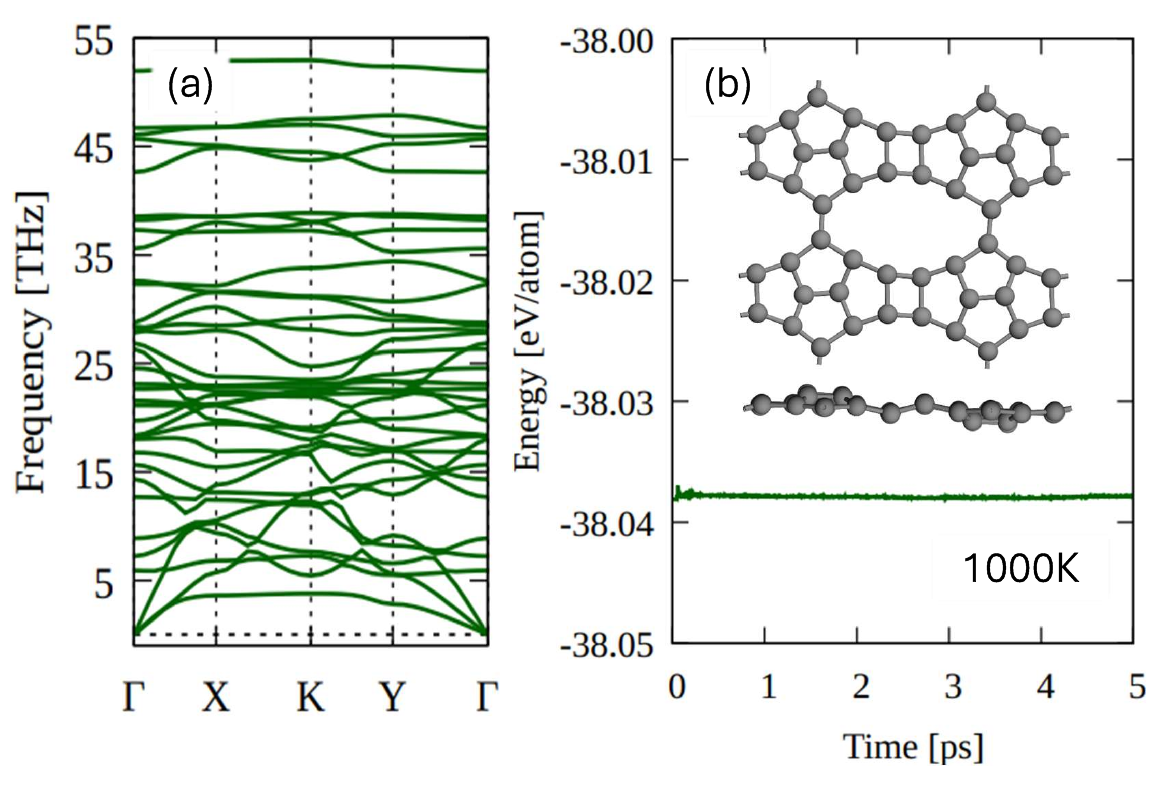}
    \caption{(a) AG phonon dispersion along the high-symmetry path $\Gamma$–X–K–Y–$\Gamma$. (b) Total energy per atom during a \SI{5}{\ps} at \SI{1000}{\kelvin}.}
    \label{fig:phonons-aimd}
\end{figure}

The AG unit cell (outlined in black line) consists of 12 carbon atoms and exhibits orthorhombic symmetry within the PMMM (D$_2$h-1) space group. AG lattice parameters are $a = 6.62$ \si{\angstrom} and $b = 5.72$ \si{\angstrom}. Bond lengths between first-neighbor atoms range from \SIrange{1.38}{1.47}{\angstrom}, indicative of slightly localized distortions from ideal sp$^2$ hybridization due to the presence of non-hexagonal rings. Despite the angular strain created by these polygons, the entire lattice preserves a flat geometry, which contrasts with the common pentagonal-based structures like penta-graphene\cite{zhang2015penta} and penta-octa-graphene\cite{laranjeira2024novel}. A repeating structural motif, highlighted in yellow, is derived from the dehydrogenated dicyclobuta[de,ij]naphthalene precursor.

An additional layer of artistic inspiration emerges from the right panel of Figure~\ref{fig:system}, which depicts a portion of a mosaic work of Athos Bulcão\cite{wanderley2006azulejo}. The characteristic interplay of symmetry and asymmetry in Bulcão's geometric compositions influenced the conceptual AG design. In particular, the mosaic's aesthetic juxtaposition of modules and voids was a visual analog for introducing ring diversity and porosity within the carbon lattice.

AG can be meaningfully compared to several other non-graphene 2D carbon allotropes. Phagraphene \cite{wang2015phagraphene}, for instance, incorporates 5-, 6-, and 7-membered rings in a distorted yet energetically favorable pattern, whereas AG achieves comparable stability while including 12-membered rings that significantly increase porosity. Popgraphene \cite{wang2018popgraphene}, with its alternating 5-8-5 and ring motifs, also lacks the same degree of geometric porosity observed in AG. The recently experimentally realized biphenylene network \cite{fan2021biphenylene} exhibits alternating 4-, 6-, and 8-membered rings, but does not match AG's ring diversity, porosity, or unique aesthetic structure. 

\begin{figure*}[!htb]
    \centering
    \includegraphics[width=\linewidth]{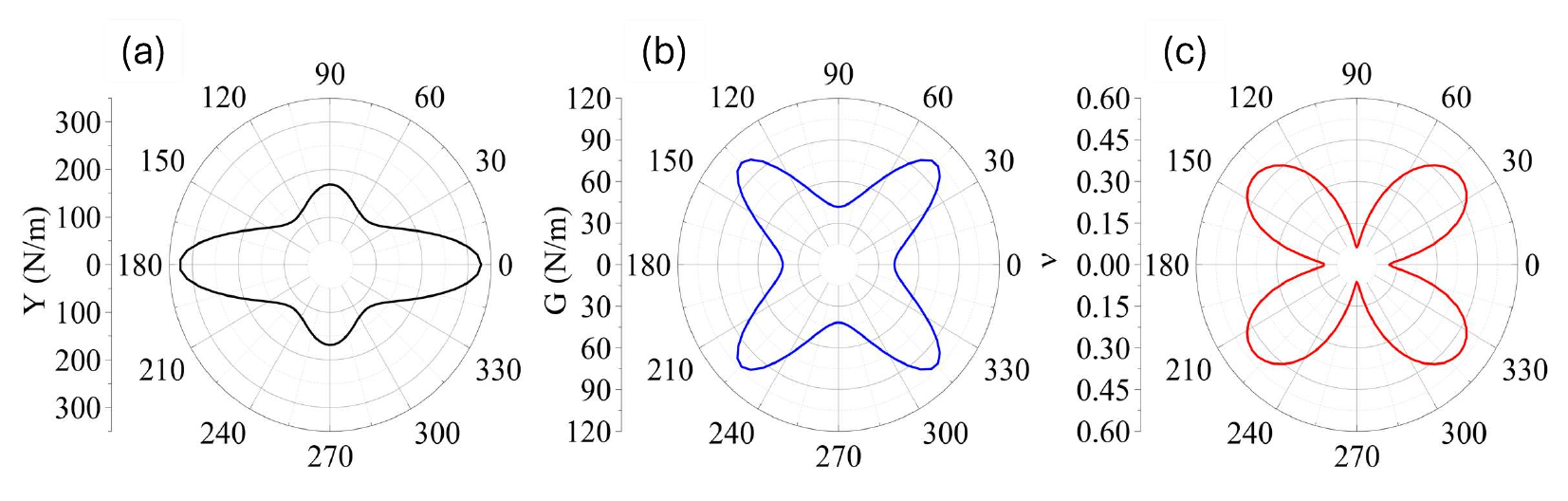}
    \caption{Polar plots of AG (a) Young's modulus ($Y(\theta)$), (b) shear modulus ($G(\theta)$), and (c) Poisson's ratio ($\nu(\theta)$). The anisotropic profiles reflect the orthorhombic symmetry and the lattice asymmetric ring connectivity.}
    \label{fig:mechprop}
\end{figure*}

The dynamical and thermal AG stability were evaluated through phonon dispersion and AIMD simulations. Fig.~\ref{fig:phonons-aimd}(a) presents the phonon band structure across the high-symmetry path $\Gamma$–X–K–Y–$\Gamma$ of the Brillouin zone. The complete absence of imaginary frequencies confirms the dynamical structural stability, indicating that the obtained optimized AG geometry corresponds to an actual local minimum on the potential energy surface. All phonon branches remain strictly positive, including the acoustic modes near the $\Gamma$ point, suggesting that no lattice instabilities arise from long-wavelength perturbations.

In addition to the vibrational analyses, the thermal stability was also evaluated using AIMD simulations at \SI{1000}{\kelvin}, as shown in Figure~\ref{fig:phonons-aimd}(b). The total energy per atom remains stable throughout the \SI{5}{\ps} simulation run time, exhibiting negligible fluctuations around a mean value of approximately \SI{-38.04}{\electronvolt/atom}. The snapshots taken at the end of the simulation (inset) reveal that the atomic configuration is preserved, with no bond breaking, reconstruction, or significant out-of-plane distortions. These results confirm that AG can maintain its structural integrity under significant thermal stress and is thus suitable for high-temperature applications.

The AG mechanical behavior was investigated through the in-plane elastic constants and the corresponding angular-dependent elastic moduli. The estimated values for the elastic constants, $C_{11} = 327.402$ \si{N/m}, $C_{22} = 169.344$ \si{N/m}, $C_{12} = 20.221$ \si{N/m}, and $C_{44} = 41.235$ \si{N/m}, satisfy the Born–Huang mechanical stability criteria for orthorhombic crystals \cite{PhysRevB.90.224104,doi:10.1021/acs.jpcc.9b09593}: $C_{11}C_{22} - C_{12}^2 > 0$ and $C_{44} > 0$, which ensures that the structure is mechanically stable against both axial and shear deformations. These conditions are satisfied, which confirms that the AG is robust from a mechanical point of view.

Figure~\ref{fig:mechprop} illustrates the directional dependence of the in-plane Young's modulus $Y(\theta)$, shear modulus $G(\theta)$, and Poisson's ratio $\nu(\theta)$, represented in polar plots. The Young's modulus (Fig.~\ref{fig:mechprop}(a)) exhibits a significantanisotropy, with maximum stiffness along the $x$-direction \SI{324.988}{N/m} and a considerably lower value along the $y$-direction \SI{120.806}{N/m}. This anisotropy arises from the non-equivalent bonding environments introduced by the asymmetric ring distribution in the unit cell.

A similar trend is observed for the shear modulus in Fig.~\ref{fig:mechprop}(b), which varies significantly with direction, further highlighting the anisotropic elastic response. The values of $G(\theta)$ vary from approximately \SI{41.235}{N/m} to \SI{102.449}{N/m} depending on the orientation, reflecting a four-fold symmetry characteristic of the orthorhombic lattice. This moderate anisotropy in the shear response suggests that the material can accommodate tangential deformations more efficiently along specific directions, which may be advantageous in flexible or strain-tunable device architectures. The relatively high shear modulus values also indicate strong angular resistance to deformation, consistent with the rigid nature of the sp$^2$ carbon network.

The polar plot for Poisson's ratio in Figure~\ref{fig:mechprop}(c) also reveals directional variations, with values ranging from approximately \SI{0.062}{} to \SI{0.499}{} depending on the loading direction, suggesting varying degrees of transverse contraction depending on the orientation of applied stress. This anisotropy is typical in low-symmetry 2D carbon systems and has been observed in other 2D carbon allotropes \cite{zhao2013mechanical,cheng2020optical,morresi2020structural,sui2017morphology}. However, AG combines this directional behavior with a high in-plane stiffness along its principal axis, making it suitable for applications requiring structural resilience and flexibility.

\subsection{Electronic and optical properties}

\begin{figure}[!htb]
    \centering
    \includegraphics[width=\linewidth]{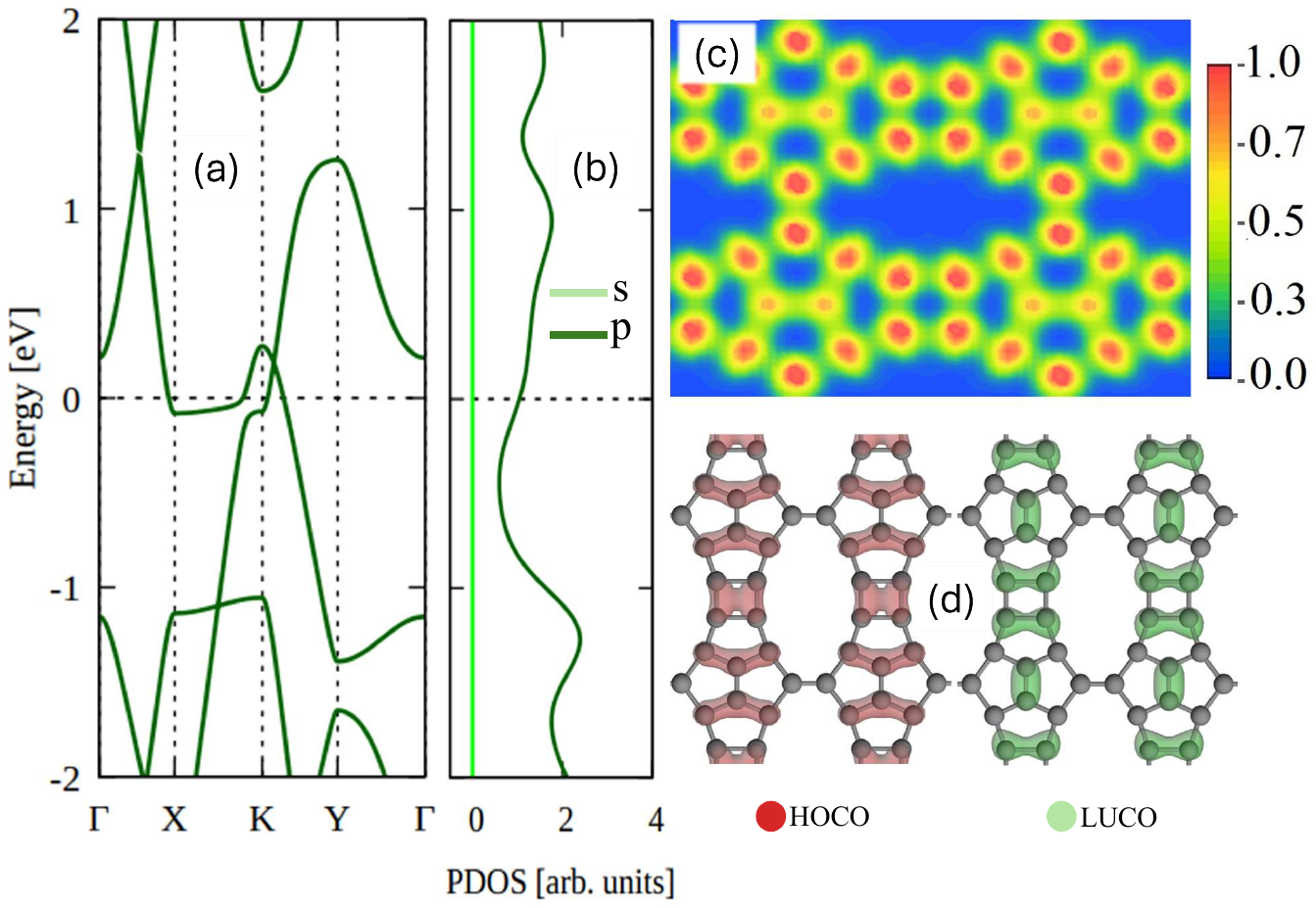}
    \caption{(a) AG electronic band structure, (b) the corresponding projected density of states (PDOS), and (c) electron localization function (ELF). (d) visualization of the HOCO (red) and LUCO (green). See text for discussions.}
   \label{fig:elecprop}
\end{figure}

The AG electronic band structure, shown in Figure~\ref{fig:elecprop}(a), reveals a metallic character with two bands crossing at the Fermi level. These bands are partially occupied and give rise to multiple cone-shaped dispersion features in both the valence and conduction regions, particularly along the $\Gamma \rightarrow Y$, $X \rightarrow K$, and $K \rightarrow Y$ paths. Such linear, cone-like dispersions are reminiscent of Dirac-like behavior, indicating the presence of massless or nearly massless quasiparticles that can exhibit high mobility. The presence of these features along multiple high-symmetry directions suggests an anisotropic yet robust transport behavior, which may be advantageous for electronic applications that require directional conductivity.

The projected density of states (PDOS), shown in Figure~\ref{fig:elecprop}(b), indicates that the electronic states near the Fermi level are dominated by $p$ orbitals, as expected for a planar sp$^2$ carbon framework. The $s$ orbital contribution is negligible throughout the entire energy range, confirming that the $\pi$ and $\pi^*$ bands, arising from the $p_z$ orbitals, govern the electronic behavior.

To further investigate the nature of the electronic distribution and bonding in AG, the electron localization function (ELF) is presented in Fig.~\ref{fig:elecprop}(c). The ELF provides insight into the degree of electron pairing and localization within the structure. High ELF values (close to 1, shown in red) are observed along the \ce{C}–\ce{C} bonds, indicating strong covalent interactions characteristic of sp$^2$ hybridization. The uniform distribution of high ELF regions across different ring sizes, including distorted four-, five-, and twelve-membered rings, confirms the persistence of localized $\sigma$-bonds despite geometric irregularities. In contrast, the interstitial regions between rings exhibit low ELF values (blue), denoting delocalized $\pi$-electron clouds responsible for the metallic character observed from the electronic band structure. This ELF profile supports strong in-plane covalent bonding with out-of-plane delocalized electronic states, which is essential for structural integrity and electrical conductivity.

Fig.~\ref{fig:elecprop}(d) depicts the highest occupied crystal orbital (HOCO) and the lowest unoccupied crystal orbital (LUCO). The HOCO is primarily localized around the 5-membered rings, while the LUCO shows delocalization across the 4 and 5-membered rings. The relevance of these characteristics for the optical transitions and excitonic effects is discussed below.

\begin{figure}[!htb]
    \centering
    \includegraphics[width=\linewidth]{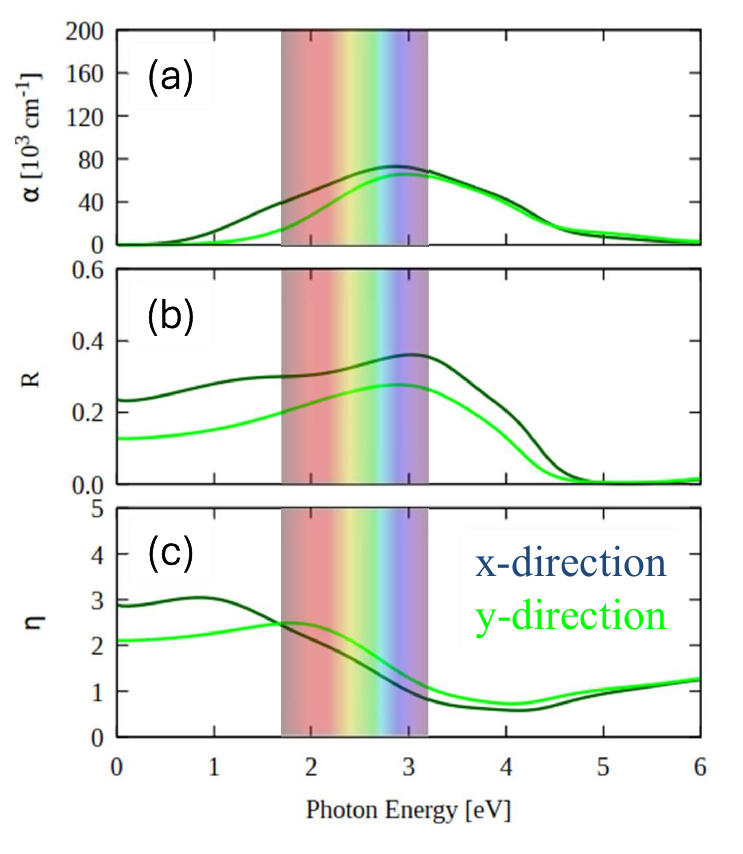}
    \caption{AG optical properties for light polarized along the $x$-direction (dark green) and $y$-direction (light green): (a) absorption coefficient $\alpha$, (b) reflectivity $R$, and (c) extinction coefficient $\eta.$}
    \label{fig:optical}
\end{figure}

The optical response of AG was investigated through its frequency-dependent dielectric function \cite{lima2023dft}, from which key optical quantities were derived. Fig.~\ref{fig:optical} displays the optical absorption coefficient $\alpha$, reflectivity $R$, and the extinction coefficient $\eta$ as functions of photon energy values, computed for light polarized along the $x$- (dark green) and $y$-directions (light green).

As shown in Fig.~\ref{fig:optical}(a), AG exhibits strong and anisotropic absorption across the visible and ultraviolet (UV) spectral regions. The onset of absorption occurs around \SI{0.6}{\electronvolt} and is followed by prominent absorption peaks in the \SI{6}{}–\SI{16}{\electronvolt} range. The absorption is generally more intense along the $x$-direction in the \SIrange{10}{14}{\electronvolt} region, suggesting a directional dependence of the optical transitions for higher photon excitations; the opposite occurs in the \SIrange{14}{20} {\electronvolt} region, where the $y$-direction has a higher absorption coefficient. This anisotropy reflects the underlying asymmetry in the atomic lattice and electronic structure induced by the non-hexagonal ring distribution.

Reflectivity results are presented in Fig.~\ref{fig:optical}(b). AG shows a reflectivity throughout the visible spectrum below \SI{40}{\percent}, being slightly higher for $x$-direction, for higher photon excitations, these values are smaller. Slight differences between the $x$ and $y$ polarization components are observed, again confirming the anisotropic optical behavior.

Fig.~\ref{fig:optical}(c) presents the extinction coefficient $\eta$, which describes the attenuation of electromagnetic waves as they propagate through the material. The extinction coefficient follows the opposite trend observed in the absorption spectrum: increased values are observed in regions where absorption is weak. This correlation is expected because both quantities are directly related to the imaginary part of the complex refractive index.

\subsection{\ce{Li}-ion diffusion on monolayer and bilayer AG}

\begin{figure*}[!htb]
    \centering
    \includegraphics[width=\linewidth]{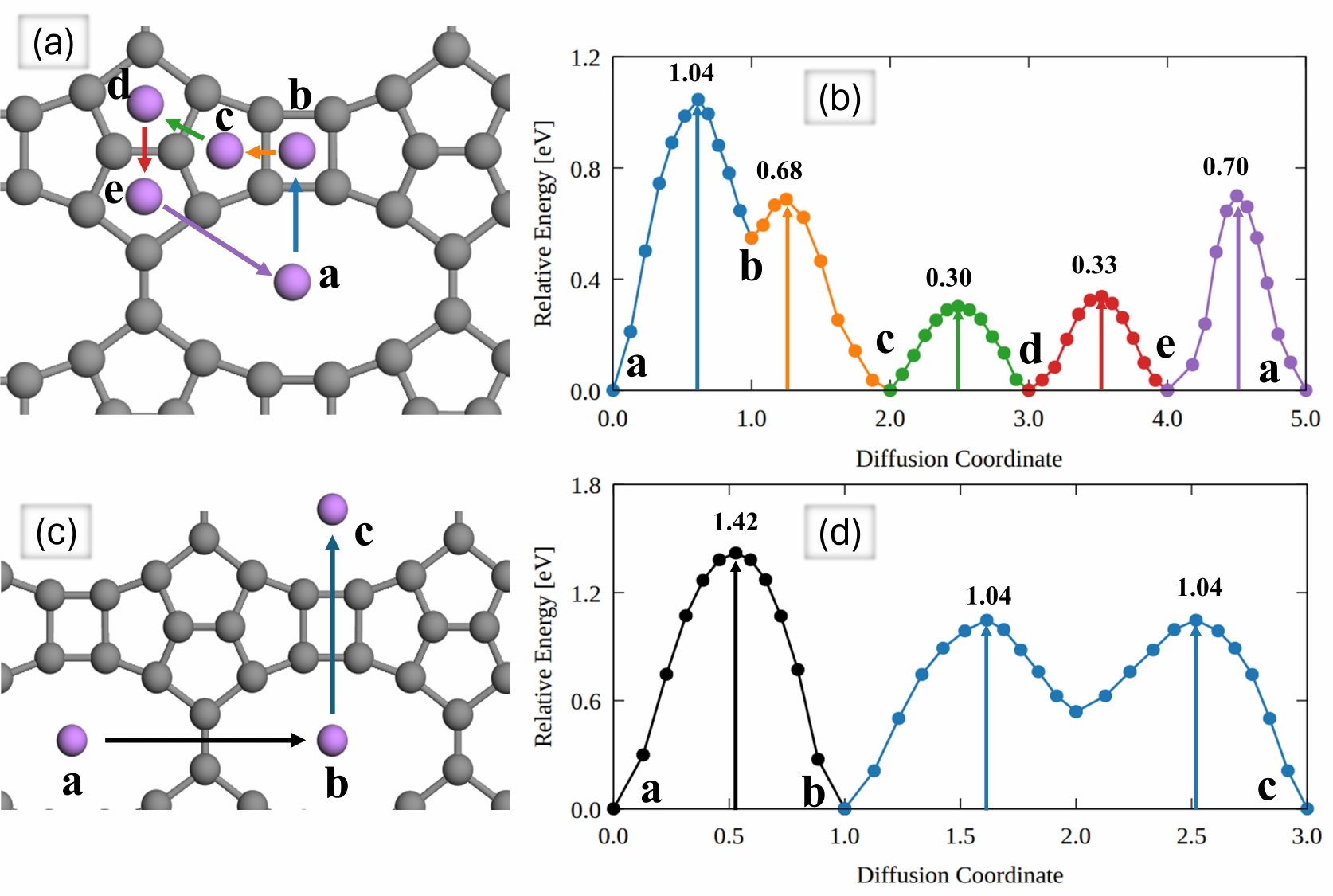}
    \caption{\ce{Li}-ion diffusion pathways on AG. (a) Top view of monolayer AG showing multiple hopping sites labeled from $a$ to $e$; (b) the corresponding energy profile for \ce{Li}-ion migration across hollow sites of the monolayer. (c) Alternative diffusion pathway through a large 12-membered ring; (d) energy barrier values associated with the bilayer model along the indicated trajectories.}
    \label{fig:diffusion_mono}
\end{figure*}

The diffusion dynamics of \ce{Li} ions across the anode material critically influence lithium-ion batteries' charge/discharge rate and overall performance. To evaluate the diffusion kinetics on AG, we considered \ce{Li}-ion migration along various hollow adsorption sites defined within its unique ring topology, as depicted in Fig.~\ref{fig:diffusion_mono}.

In the monolayer configuration, several migration pathways were analyzed between high-symmetry hollow sites (Fig.~\ref{fig:diffusion_mono}(a)). The energy profile in Fig.~\ref{fig:diffusion_mono}(b) shows that diffusion barriers vary substantially depending on the ion's specific path. The most restrictive path corresponds to the $a \rightarrow b$ transition, which crosses a cyclobutadiene-like region and results in a barrier of \SI{1.04}{\electronvolt}. This high barrier is attributed to the strained bonding environment and localized electronic states within the four-membered ring.

In contrast, diffusion along the $c \rightarrow d$ and $d \rightarrow e$ directions is significantly more favorable, with energy barriers of only \SI{0.30}{\electronvolt} and \SI{0.33}{\electronvolt}, respectively. These paths involve motion across pentagonal motifs, suggesting that five-membered rings in AG facilitate \ce{Li} migration by reducing topological constraints. Overall, the minimum diffusion barrier of \SI{0.30}{\electronvolt} highlights AG's potential for efficient ion transport, outperforming established 2D materials such as graphene (\SI{0.32}{\electronvolt})~\cite{uthaisar2010edge} and phosphorene (\SI{0.76}{\electronvolt})~\cite{zhao2014potential}, and comparable to emerging candidates like Petal-Graphyne (\SI{0.28}{\electronvolt} )~\cite{lima2025petal}, Irida-Graphene (\SI{0.19}{\electronvolt})~\cite{xiong2024theoretical}, biphenylene network (\SI{0.23}{\electronvolt})~\cite{duhan20232}, PolyPyGY~\cite{LIMA2025116099}, and T-graphene (\SI{0.37}{\electronvolt})~\cite{hu2020theoretical}.

A different scenario is observed when considering diffusion across the center of the larger 12-membered rings, as modeled in Fig.~\ref{fig:diffusion_mono}(c). As shown in Fig.~\ref{fig:diffusion_mono}(d), energy barriers increase significantly in this configuration, ranging from \SIrange{1.04}{1.42}{\electronvolt}. This result suggests that the extended inter-ring covalent bonds in the dodecagonal major ring hinder \ce{Li}-ion mobility, corroborating the orbital localization observed in the HOCO–LUCO distributions (see Fig.~\ref{fig:elecprop}(d)), where electronic density accumulates near the atomic centers.

To quantify the influence of temperature on \ce{Li}-ion mobility, we computed the diffusion coefficients ($D_{\text{coeff}}$) using the Arrhenius-type relation:

\begin{equation}
D_{\text{coeff}}(T) = L^2 \nu_0 \exp \left( -\frac{\Delta E_b}{k_B T} \right),
\end{equation}

\noindent where $\Delta E_b$ is the energy barrier for diffusion, $L$ is the hopping distance, $\nu_0$ is the attempt frequency (assumed to be \SI{10}{\tera\hertz}), $T$ is the absolute temperature, and $k_B = 8.62 \times 10^{-5}$ \si{eV/K} is the Boltzmann constant \cite{gao2023twin}.

\begin{figure}[!htb]
    \centering
    \includegraphics[width=\linewidth]{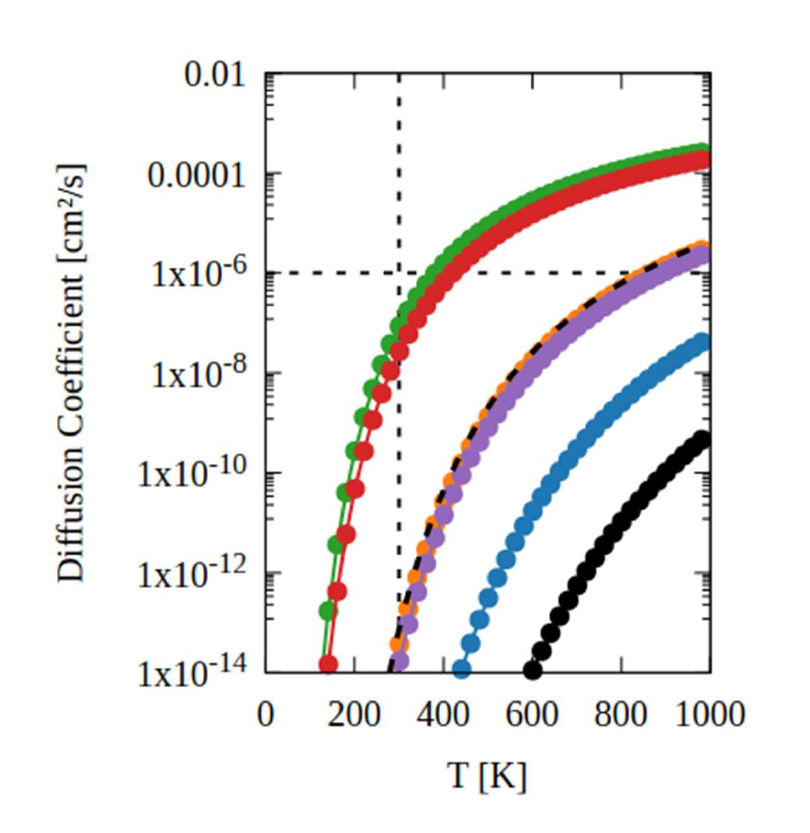}
    \caption{Temperature-dependent diffusion coefficients ($D_{\text{coeff}}$) for \ce{Li}-ion migration across different pathways in monolayer and bilayer AG. The dashed line at \SI{300}{\kelvin} refers to the \ce{Li}-ion diffusion on the graphene surface for comparison.}
\label{fig:diffusion_coeff}
\end{figure}

Fig.~\ref{fig:diffusion_coeff} presents the diffusion coefficients as a function of temperature for various migration pathways in monolayer and bilayer AG. As expected, all curves follow an exponential trend, with increasing diffusion rates at higher temperatures. The lowest-energy monolayer pathways (barriers of \SI{0.30}{\electronvolt} and \SI{0.33}{\electronvolt}) result in diffusion coefficients exceeding $10^{-6}$ \si{\square\cm/s} at room temperature, placing AG as a promising 2D material for high-rate lithium-ion transport. These values surpass those of conventional anode materials and are comparable to or better than other novel carbon allotropes, such as graphene \cite{uthaisar2010edge}, T-graphene \cite{hu2020theoretical}, biphenylene network \cite{duhan20232}, and Irida-Graphene \cite{xiong2024theoretical}.

\begin{figure*}[!htb]
    \centering
    \includegraphics[width=\linewidth]{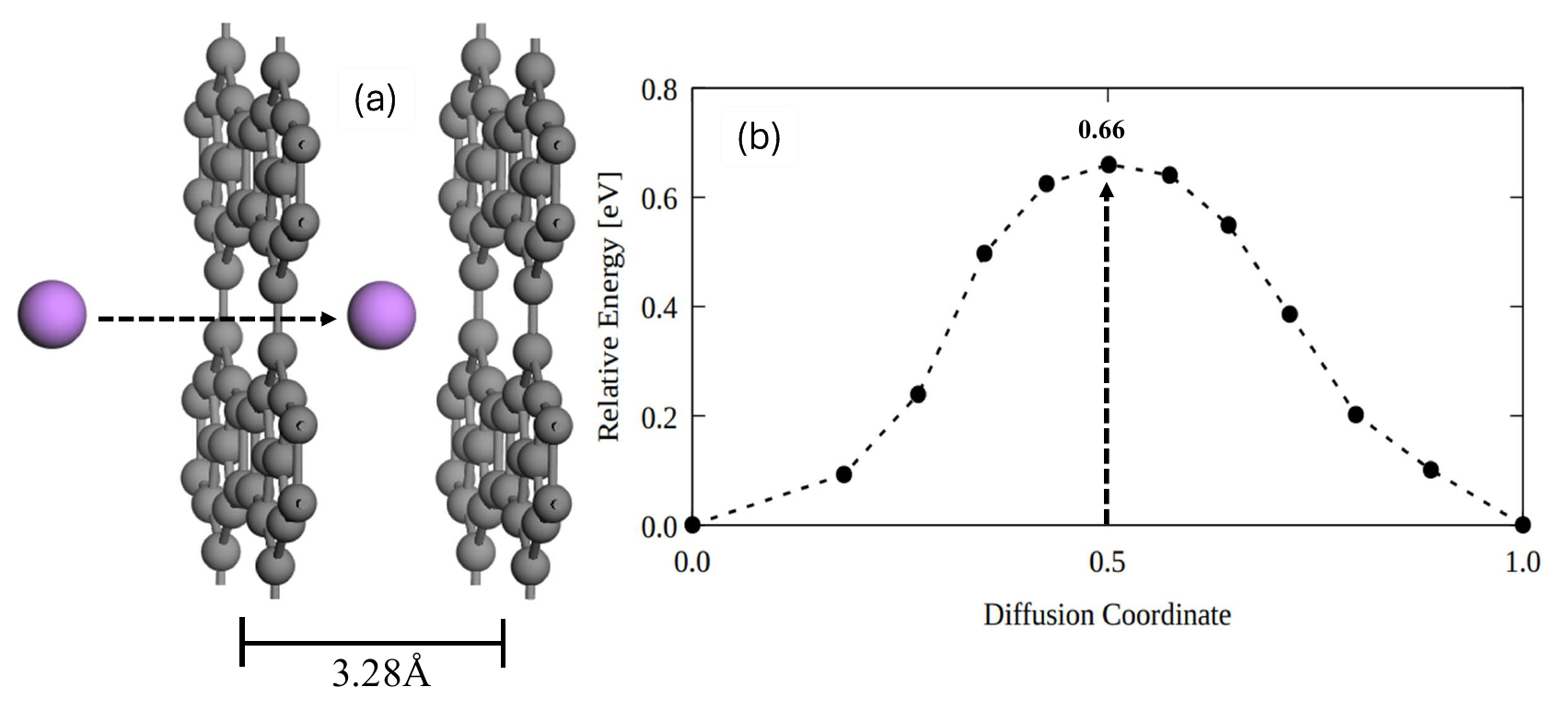}
    \caption{Interlayer \ce{Li}-ion diffusion in AA-stacked bilayer AG. (a) \ce{Li} migration path between two adjacent AG layers separated by \SI{3.28}{\angstrom}. (b) The energy profile along the interlayer diffusion coordinate shows a maximum energy barrier of \SI{0.66}{\electronvolt} at the midpoint.}
    \label{fig:interlayer_diff}
\end{figure*}

To investigate the interlayer diffusion properties of AG, we modeled an AA-stacked bilayer configuration, as illustrated in Fig.~\ref{fig:interlayer_diff}(a). A single \ce{Li}-ion was initially placed in the center of a 12-membered ring on the surface of the bottom AG sheet and allowed to migrate perpendicularly along the $z$-direction toward the corresponding site on the top layer. The interlayer spacing is approximately \SI{3.28}{\angstrom}, consistent with typical van der Waals gaps in carbon-based layered materials \cite{rathanasamy2021carbon}.

The calculated diffusion energy profile, shown in Fig.~\ref{fig:interlayer_diff}(b), reveals a symmetric potential energy landscape with a maximum energy barrier of \SI{0.66}{\electronvolt} at the midpoint between the two layers. This barrier is substantially higher than the lowest in-plane diffusion pathways (\SI{0.30}{}–\SI{0.33}{\electronvolt}), indicating that out-of-plane migration is less favorable from a kinetic point of view. The increase in barrier height is mainly attributed to strong $\pi$–$\pi$ stacking interactions between the layers and the lack of bonding sites in the interstitial region, forcing the \ce{Li}-ion to traverse a less electronically favorable path.

\subsection{Storage capacity and OCV range}

The \ce{Li} adsorption characteristics of AG were explored as a function of the number of adsorbed atoms per supercell. The adsorption energy for each configuration was calculated using the equation: 
\begin{equation}
E_{\text{ads}} = \frac{E_{\text{N--Li+AG}} - \left( N E_{\text{Li}} + E_{\text{AG}} \right)}{N}
\end{equation}

\noindent where $E_{\text{N--Li+AG}}$ is the total energy of the AG supercell with $N$ adsorbed \ce{Li} atoms, $E_{\text{AG}}$ is the energy of pristine AG, and $E_{\text{Li}}$ is the energy of an isolated \ce{Li} atom. 

\begin{figure}
    \centering
    \includegraphics[width=\linewidth]{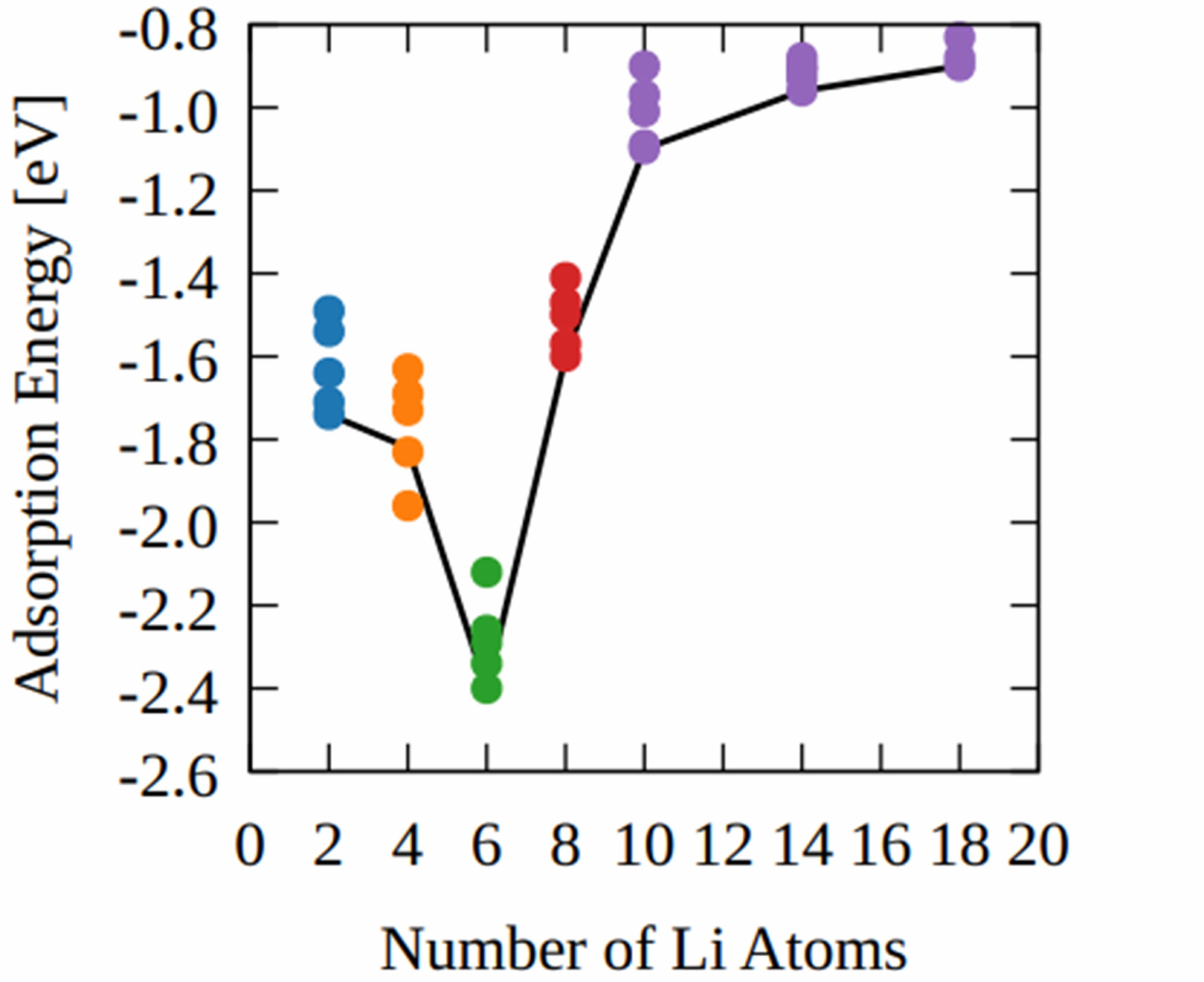}
    \caption{Adsorption energy as a function of the number of \ce{Li} adsorbed atoms on AG.}
    \label{fig:adsorption}
\end{figure}

In this study, we consider double-sided symmetrical adsorption to gain a more comprehensive understanding of the theoretical capacity of AG. As shown in Fig.~\ref{fig:adsorption}, the adsorption energy remains negative for all configurations, indicating that lithium binding is spontaneous and favorable. This observation is critical for battery safety, as positive values of \( E_{\text{ads}} \) could suggest dendrite nucleation. Furthermore, it can be observed that until reaching \( N_{\text{opt}} = 6 \), the adsorption energy becomes progressively more negative. After this point, it gradually decreases as the maximum storage capacity of 18 \ce{Li} atoms is approached. A possible explanation for this trend is that, initially, the interaction distance between the first adsorbed \ce{Li} atoms remains relatively large. As more \ce{Li} atoms are adsorbed at the atomic sites of the structure, repulsive interactions increase, leading to a mitigation of \( E_{\text{ads}} \), resulting in an energy range of \SIrange{-2.39}{-0.9}{\electronvolt}, consistent with the literature\cite{lima2025petal, ullah2024theoretical, wang2019planar}. As discussed in the previous section, the mobility of \ce{Li} on the AG monolayer is high, providing a favorable condition for charge transfer modulation between the \ce{Li}-\ce{Li} metals and \ce{Li} to the AG structure. 

Based on the maximum lithiation level ($N_{\text{max}} = 18$), the theoretical storage capacity of AG can be calculated using the relation: 

\begin{equation}
C = \frac{N_{\text{max}} F}{M},    
\end{equation}

\noindent where $F$ is the Faraday constant (96485.332 A$\times$s/mol), and $M$ is the molar mass of the fully relaxed AG supercell. This yields a high theoretical capacity of \SI{836.78}{mAh/g}, which surpasses conventional graphite (\SI{372}{ mAh/g}) and graphene (\SI{744}{mAh/g}) electrodes \cite{wang2009sn,chen2023new}. However, this capacity is comparable to or even exceeds that of recently predicted 2D carbon allotropes for \ce{Li} anodes, such as tolanene (\SI{638}{mAh/g}) \cite{ullah2024theoretical}, penta-graphyne (\SI{680}{mAh/g}) \cite{deb2022two}, petal-graphyne (\SI{1004}{mAh/g}) \cite{lima2025petal}, and holey-graphyne (\SI{873}{mAh/g}) \cite{sajjad2023two}.

\begin{figure}[!htb]
    \centering
    \includegraphics[width=\linewidth]{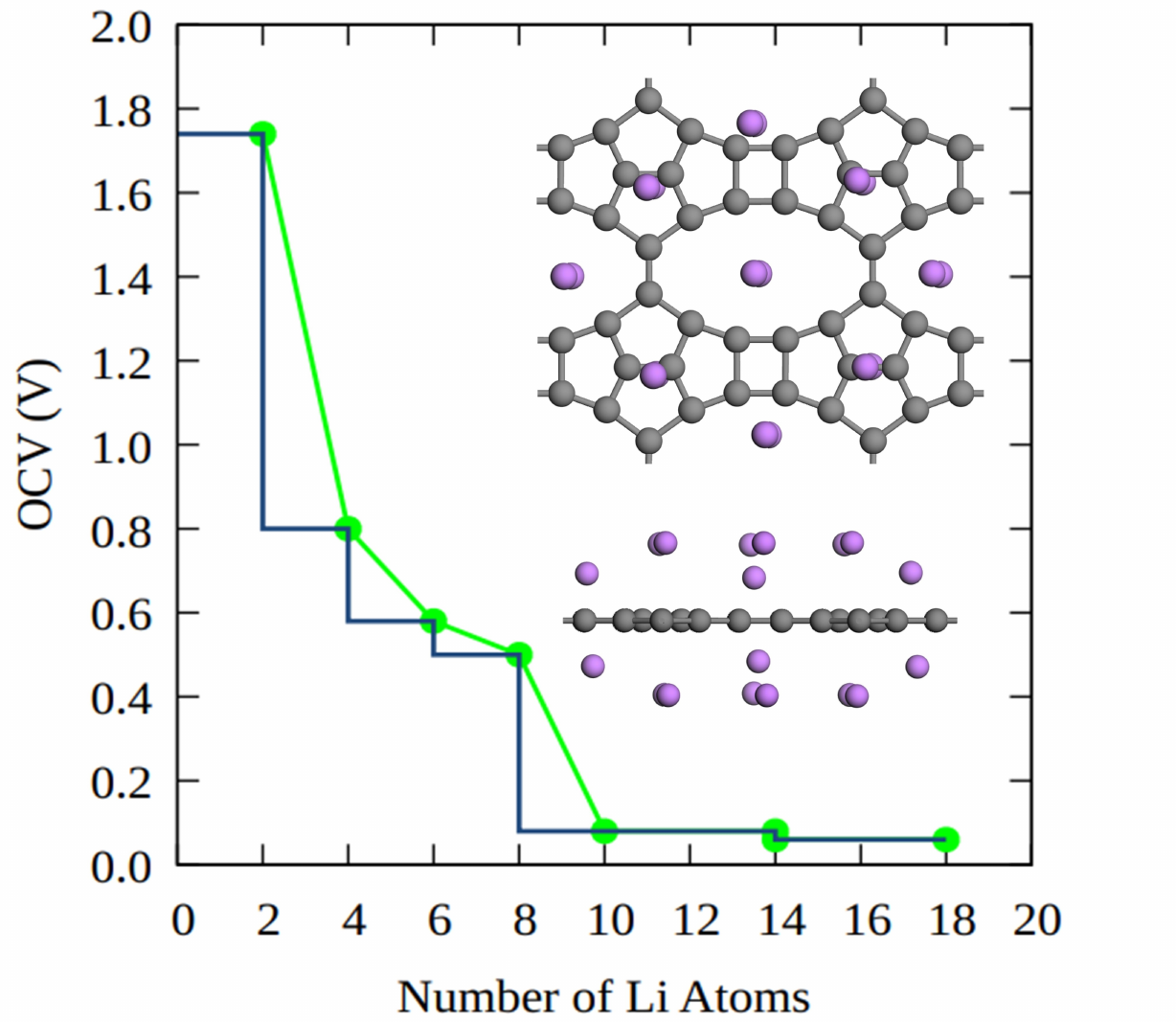}
    \caption{Open-circuit voltage (OCV) profile as a function of \ce{Li} concentration. The average OCV is \SI{0.54}{\volt}. The inset panels illustrate the top and side views of the fully lithiated AG structure.}
    \label{fig:ocv}
\end{figure}

To evaluate the electrochemical performance of AG as an anode material, the open-circuit voltage (OCV) was computed using the following equation:

\begin{equation}
\text{OCV} = \frac{E_{\text{AG}} + N E_{\text{Li}} - E_{\text{N--Li+AG}}}{N e}
\end{equation}

\noindent where $e$ is the elementary charge. Fig.~\ref{fig:ocv} shows the OCV as a function of the number of adsorbed \ce{Li} atoms. The OCV profile exhibits a staircase-like behavior, with an initial high value of \SI{1.84}{\volt} at low coverage, followed by a steady decrease as additional \ce{Li} atoms occupy energetically less favorable sites. The average OCV is approximately \SI{0.54}{\volt}, remaining positive throughout the lithiation range, ensuring the prevention of \ce{Li}-metal plating and supporting good cycling performance. The average OCV value (\SI{0.54}{\volt}) is within the ideal range (typically \SI{0.1}{}–\SI{1.0}{\volt}) for LIB anodes \cite{kirklin2013high}.

The inset of Fig.~\ref{fig:ocv} shows the optimized fully lithiated configuration, where \ce{Li} atoms are uniformly adsorbed on both sides of the AG monolayer. The minimum \ce{Li}–\ce{Li} distance remains larger than \SI{2.8}{\angstrom}, effectively avoiding clustering and favoring the homogeneous ion distribution.

\section{Conclusion}

In summary, we proposed Athos-Graphene (AG), a novel two-dimensional carbon allotrope. Through DFT calculations, we demonstrated that AG exhibits a unique porous and planar structure composed of four-, five-, and twelve-membered carbon rings arranged in an orthorhombic lattice. The material is dynamically, thermally, and mechanically stable, as evidenced by its phonon dispersion, AIMD simulations at \SI{1000}{\kelvin}, and compliance with Born-Huang criteria. AG also shows mechanical anisotropy, with high Young’s and shear moduli along the principal crystallographic directions, supporting its robustness and flexibility.

Electronic structure calculations revealed a metallic character with significant contribution from delocalized $\pi$-electrons, while optical analyses showed anisotropic absorption in the visible and ultraviolet regions, reinforcing AG’s multifunctionality. AG shows excellent performance as a lithium-ion battery anode, with strong adsorption energies, low diffusion barriers (as low as \SI{0.30}{\electronvolt}), and a high theoretical storage capacity of \SI{836.78}{mAh/g}. The open-circuit voltage remains low and positive throughout the lithiation range, with an average value of \SI{0.54}{\volt}. These combined properties make AG a promising candidate as an anodic material for high-performance lithium-ion batteries.

\section*{Data access statement}
Data supporting the results can be accessed by contacting the corresponding author.

\section*{Conflicts of interest}
The authors declare no conflict of interest.

\section*{Acknowledgements}
This work was supported by the Brazilian funding agencies Fundação de Amparo à Pesquisa do Estado de São Paulo - FAPESP (grant no. 2022/03959-6, 2022/14576-0, 2020/01144-0, 2024/05087-1, and 2022/16509-9), and National Council for Scientific, Technological Development - CNPq (grant no. 307213/2021–8). A.C.D. acknowledges the financial support from FAP-DF grants $00193-00001817/2023-43$ and $00193-00002073/2023-84$, CNPq grants $408144/2022-0$, $305174/2023-1$, $444069/2024-0$ and $444431/2024-1$; and also the computational resources provided by Cenapad-SP (projects 897) and Lobo Carneiro HPC (project 133) are also greatly appreciated. A.C.D. and L.A.R.J. acknowledges the financial support PDPG-FAPDF-CAPES Centro-Oeste $00193-00000867/2024-94$. L.A.R.J. acknowledges the financial support from FAP-DF grants $00193.00001808$ $/2022-71$ and $00193-00001857/2023-95$, FAPDF-PRONEM grant $00193.00001247/2021-20$, PDPG-FAPDF-CAPES Centro-Oeste $00193-00000867/2024-94$, and CNPq grants $350176/2022-1$ and $167745/2023-9$. 

\printcredits

\bibliography{cas-refs}

\end{document}